\documentclass[
aps,
prb,
twocolumn,
amsfonts,
amsmath,
amssymb,
mathtools,
superscriptaddress,
longbibliography,
noeprint,
nofootinbib,
10pt
]{revtex4-2}

\usepackage{graphicx}
\usepackage{bm}
\usepackage[dvipsnames]{xcolor}
\usepackage[english]{babel}
\usepackage{dsfont}
\usepackage{comment}
\usepackage[caption=false]{subfig}
\usepackage[shortlabels]{enumitem}
\usepackage[colorlinks,%
            linkcolor=BrickRed,%
            citecolor=MidnightBlue,%
            urlcolor=MidnightBlue]{hyperref}

\newcounter{ls}

\newcommand{\bc}{\begin{center}}
\newcommand{\ec}{\end{center}}
\def\ba#1{\begin{array}{#1}\displaystyle}
\newcommand{\ea}{\end{array}}

\newcommand{\beq}{\begin{equation}}
\newcommand{\eeq}{\end{equation}}
\newcommand{\beqa}{\begin{eqnarray}}
\newcommand{\eeqa}{\end{eqnarray}}

\newcommand{\n}{\nonumber\\}
\newcommand{\bi}{\begin{itemize}}
\newcommand{\ei}{\end{itemize}}

\newcommand{\Tr}{{\rm Tr}}

\newcommand{\varep}{\varepsilon}

\newcommand{\ii}{{\rm i}}

\begin{document}

\title{Dynamics of Loschmidt echoes from operator growth\\in noisy quantum many-body systems}

\author{Takato Yoshimura}
\email{takato.yoshimura@physics.ox.ac.uk}
\affiliation{All Souls College, Oxford OX1 4AL, U.K.}
\affiliation{Rudolf Peierls Centre for Theoretical Physics, University of Oxford, 1 Keble Road, Oxford OX1 3NP, U.K.}

\author{Lucas S\'a}
\email{ld710@cam.ac.uk}
\affiliation{TCM Group, Cavendish Laboratory, Ray Dolby Centre, University of Cambridge, JJ Thomson Avenue, Cambridge, CB3 0US UK}

\begin{abstract}    
We study the dynamics of Loschmidt echoes in noisy quantum many-body systems without conservation laws. We first show that the operator Loschmidt echo in noisy unitary dynamics is equivalent to the operator norm of the corresponding dissipative dynamics upon noise averaging. We then analyze this quantity in two complementary ways, revealing universal dynamical behavior. First, we develop a heuristic picture for generic Floquet systems that connects Loschmidt echoes, out-of-time-order correlators, and operator growth, which is valid at any dissipation strength. We assert that the Loschmidt echo has two dynamical regimes depending on the time $t$ and the strength of the noise $p$: Gaussian decay for $pt\ll1$ and exponential decay (with a noise-independent decay rate) for $pt\gg1$. Lastly, we rigorously prove all our results for a solvable chaotic many-body quantum circuit, the dissipative random phase model---thus providing exact insight into dissipative quantum chaos.
\end{abstract}

\maketitle

\section{Introduction}

Current and future applications of quantum information processing platforms require an accurate understanding and control of the flow and storage of quantum information in many-body systems. In chaotic quantum many-body systems, information is quickly spread across all degrees of freedom, with initially local operators growing and becoming supported in large spatial regions~\cite{roberts2015JHEP,Keyserlingk_Operator_2018,Nahum_Operator_2018,Khemani_Diffusive_2018,parker2019PRX,nandy2015PhysRep}. However, environmental noise poses a significant obstacle to this picture: as the system becomes entangled with external degrees of freedom, averaged or reduced density matrices undergo decoherence in complex, uncontrollable ways. Such decoherence greatly influences the nature of dynamics, giving rise to operator-growth behaviours in open systems that are qualitatively different from those in closed systems~\cite{zhang2019PRB,yoshida2019PRX,touil2021PRXQ,liu2023PRR,schuster2023PRL,bhattacharya2022JHEP,bhattacharjee2023JHEP,bhattacharya2023JHEP,srivatsa2024PRB}.

A fruitful approach to this problem comes from the relation of operator growth to quantum chaos, particularly quantum uncertainty and sensitivity to perturbations of the dynamics~\cite{yan2020PRL,romero2019JHEP,garcia2022OTOC,bhattacharyya2022WEB}.
The Loschmidt echo~\cite{peres1984,jalabert2001,jacquod2001} quantifies the sensitivity of the system to small perturbations in the dynamics, defined as the overlap of two states evolved under two slightly different unitary dynamics. Depending on the chaotic or regular nature of the evolution and the strength of the perturbation, the Loschmidt echo decay takes on different functional forms. For strongly perturbed chaotic dynamics, the decay is exponential, while the decay is Gaussian in weak-perturbation regimes. Because the Loschmidt echo is theoretically and experimentally accessible (ee, e.g., the review Ref.~\cite{goussev2012arXiv} and Ref.~\cite{google2025} for a recent experimental measurement of a closely related correlation function in a quantum simulator), it has been intensively studied for over four decades; see Refs.~\cite{gorin2006,Jacquod2009,goussev2012arXiv,goussev2016PTRSA} and references therein. While most of these works focused on single-body systems, our focus will be on quantum many-body systems (see, e.g., Refs.~\cite{silva2008PRL,torres-herrera2014PRE,serbyn2017PRB,sanchez2020PRL,karch2025,carignano2025} for works in the many-body context).

The last decade has taught us that an effective strategy to tackle the apparent intractability of quantum many-body dynamics is to replace the actual microscopic dynamics with effective random evolution, over which we can average. Random quantum circuits (RUCs)~\cite{Nahum_entanglement_2017,Nahum_Operator_2018,Keyserlingk_Operator_2018,Rakovszky_Diffusive_2018,Khemani_Diffusive_2018,Nahum_entanglement_2019,RQCreview2022} and random Floquet circuits (RFCs)~\cite{Kos_ManyBody_2018,Chan_Solution_2018,Chan_Spectral_2018,Bertini_Exact_2018,Bertini_Exact_2019,Friedman_Spectral_2019,chan2019PRL,Garratt_Local_2021,chan2021PRR,yoshimura2023operator} played an instrumental role in refining our understanding of entanglement spreading, operator growth, spectral statistics, and thermalization in generic quantum many-body systems. 

In this paper, we embrace this paradigm and compute the Loschmidt echo for dynamics that are periodic but with perturbations that are random in time. Besides providing an analytical handle for exact calculations, such perturbations also capture the essential features of the action of an external environment on the system, providing a deeper understanding of noisy Loschmidt dynamics. After averaging over the noise realizations, the Loschmidt echo of a local operator is mapped to the evolution of the operator norm under \emph{dissipative Floquet dynamics}. Exploiting the relation between the norm, the operator size, and the out-of-time-order correlators (OTOC), we provide a refined understanding of, simultaneously, noisy Loschmidt dynamics and open-system operator growth~\cite{schuster2023PRL} at \emph{arbitrary perturbation/dissipation strength}. As a crucial aspect of our work, we rigorously prove all the aspects of this heuristic approach for a particular exactly solvable dissipative many-body system, the dissipative random phase model (DRPM)~\cite{yoshimura2024,yoshimura2025}.

\vfill

\section{Loschmidt echo in noisy dynamics}

The systems we consider in this paper are one-dimensional locally-interacting Floquet circuits defined on a $q$-state spin lattice of size $L$. For simplicity, we assume that there is no conservation law in the system so that every local operator spreads in a qualitatively similar way.
The time-evolution protocol consists of a forward time evolution $U$, followed by a backward time evolution $U'$. In a noisy system, the time-reversal operation is implemented imperfectly and $U'\neq U^\dagger$. The dynamics consists of a Floquet protocol where we repeatedly apply a one-step unitary $W$, but such that, after each step, the system also applies a randomly sampled unitary $V_\tau$ (the noise). We therefore have $U=\prod_{\tau=1}^t (V_\tau W)$. When we reverse time, we repeatedly apply the gate $W^\dagger$, but since we have no control over the noise, the system still applies a randomly sampled gate $V_\tau^\dagger$ from the same distribution \emph{before} each step, $U'=\prod_{\tau=1}^t(W^\dagger V_\tau^\dagger)$. (A different protocol would consist of applying noise from the \emph{same} distribution \emph{after} each time step in the backward time evolution as well---as one can argue the noise should be agnostic as to whether we are performing forward or backward evolution. Below, we will average over the noise, after which the two protocols yield qualitatively similar results. We choose the former protocol to make contact with other quantities of interest.) 

The Loschmidt echo is the survival probability of an initial state $|\psi_0\rangle$ under the imperfect time reversal, $|\langle\psi_0|U'U|\psi_0\rangle|^2=\Tr[\varrho'(t)\varrho(t)]$, where $\varrho(t)=U\varrho_0U^\dagger$, $\varrho'(t)=U'\varrho_0U'^\dagger$, and $\varrho_0=|\psi_0\rangle\langle \psi_0|$. In the Heisenberg picture, we can also define the \emph{operator Loschmidt echo}:
\begin{equation}
    M_{\mathcal{O}}(t)=\langle \tilde{\mathcal{O}}'(t)\tilde{\mathcal{O}}(t)\rangle,
\end{equation}
where $\tilde{\mathcal{O}}(t)=U^\dagger \mathcal{O}U$ is the unitarily time-evolved operator $\mathcal{O}$, $\tilde{\mathcal{O}}'(t)= U'^\dagger\mathcal{O}U'$, and $\langle\cdots\rangle=D^{-1}\Tr[\cdots]$ is the infinite-temperature average. 

If we average over the noise, the operator Loschmidt echo becomes the operator norm of $\mathcal{O}(t)$,
\begin{equation}
    \mathbb{E}_V[M_{\mathcal{O}}(t)]=\langle \mathcal{O}(t)\mathcal{O}(t)\rangle,
\end{equation}
under \emph{dissipative} Floquet dynamics, $\mathcal{O}(t)=(\mathcal{W}^\dagger)^t(\mathcal{O})$. The the quantum channel $\mathcal{W}^\dagger(\mathcal{O})=\Phi^\dagger(W^\dagger\mathcal{O}W)$, where $\Phi^\dagger(\mathcal{O})=\int \mathrm{d} V P(V) V^\dagger\mathcal{O} V$ results from averaging over the noise with probability distribution $P(V)$.
While our results hold for any noise distribution $P(V)$, for concreteness, we will focus on the following illustrative example. We fix a Hermitian basis of operators $P_\alpha=P_{\alpha^1}\otimes\cdots\otimes P_{\alpha^L}$, where $\alpha^x=0,\dots,q^2-1$. The on-site basis operator $P_{\alpha^x}$ acting on site $x$ is normalized as $q^{-1}\Tr(P_{\alpha^x}P_{\beta^x})=\delta_{\alpha^x\beta^x}$ so that $\langle P_{\alpha}P_{\beta}\rangle=\delta_{\alpha\beta}$. We also demand that the nonidentity basis operators $P_{\alpha^x}$ be normalized such that $P_{\alpha^x}^2=I_q$ (this is possible by choosing $q=2^b$ for some integer $b$, in which case $P_{\alpha^x}$ reduces to a tensor product of Pauli matrices). The noise model consists of clean evolution (i.e., $V=I_q$) with probability $1-p(q^2-1)/q^2$ or any one of the $q^2-1$ basis operators $P_{\alpha^x}$ at site $x$ with probability $p/q^2$. After averaging, it gives rise to the depolarizing channel, $\Phi=\bigotimes_x \Phi_x$, $\Phi_x(\varrho_x)=(1-p)\varrho_x+p I_q/q$. The perturbation strength $p$ of the Loschmidt echo dynamics becomes the dissipation strength of the averaged dynamics.

\section{Average operator size and OTOCs}

The operator norm $\langle \mathcal{O}(t)\mathcal{O}(t)\rangle$ also naturally arises from the study of operator growth under dissipative dynamics.
We study the average operator size $\Sigma(t)$ of a local operator $\mathcal{O}$ given by~\cite{Roberts2018,qi_2019_operatorsizegrowth,qi2019JHEP,schuster_2022_mbq_teleport,schuster2023PRL,bhattacharjee2023JHEP,mori2024} 
\begin{equation}\label{eq:operator_size}
\begin{split}
    \Sigma(t)
    &=\frac{\langle \mathcal{O}(t)\mathcal{S}\{\mathcal{O}(t)\}\rangle}{\langle \mathcal{O}(t)\mathcal{O}(t)\rangle}.
\end{split}
\end{equation}
Here, the size operator $\mathcal{S}$, defined by
\begin{equation}
    \mathcal{S}\{\mathcal{O}\}=\frac{1}{q^2}\sum_{x=1}^L\sum_{P_{\alpha^x}\in\mathcal{P}_x}\left(\mathcal{O}-P_{\alpha^x}\mathcal{O}P_{\alpha^x}\right),
\end{equation}
with $\mathcal{P}_x$ the set of the on-site basis operators at site $x$,
measures the number of nonidentity operators (i.e., $a$) in $\mathcal{O}$ if $\mathcal{O}$ is a single operator string; if $\mathcal{O}$ is a superposition of several strings, then $\mathcal{S}$ averages their sizes weighted by their respective coefficients in the superposition. 

The average operator size Eq.~\eqref{eq:operator_size} is closely related to the OTOC defined as the average square commutator
\begin{equation}\label{eq:def:OTOC}
    \mathcal{C}_{\mathcal{O}_1\mathcal{O}_2}(x,t)
    =-\frac{1}{2}\langle[\mathcal{O}_1(0,t),\mathcal{O}_2(x,0)]^2\rangle,
\end{equation}
where $\mathcal{O}_i(x,t)$ is an arbitrary operator initially supported on site $x$ and evolved to time $t$. Expanding the right-hand side of Eq.~(\ref{eq:def:OTOC}) for $\mathcal{O}_1(0,t)=\mathcal{O}(t)$ with $\langle\mathcal{O}(0)\mathcal{O}(0)\rangle=1$ and $\mathcal{O}_2(x,0)=P_{\alpha^x}$, the OTOC we shall evaluate reads
\begin{equation}\label{eq:def:OTO2}
\mathcal{C}(x,t)=\langle \mathcal{O}(t)\mathcal{O}(t)\rangle-\langle \mathcal{O}(t)P_{\alpha^x}\mathcal{O}(t)P_{\alpha^x}\rangle, 
\end{equation}
where we dropped the subscript for brevity.
It follows that the average operator size, the OTOC, and the operator norm are related via
\begin{equation}
\label{eq:OTOC_size}
    \Sigma(t)
    =\frac{1}{q^2} \sum_{x=1}^L\sum_{P_{\alpha^x}\in\mathcal{P}_x} \frac{\mathcal{C}(x,t)}{\langle \mathcal{O}(t)\mathcal{O}(t)\rangle}.
\end{equation} 

\section{Generic picture}

\subsection{Dynamics of the operator size}

The dynamics of both the operator size average and the Loschmidt echo have two competing contributions, namely, bulk dissipation and operator growth. Here we provide a heuristic argument to illustrate the generic behaviors of these quantities, starting with the former.

In the case of closed systems, operator dynamics is the only contribution to the growth of the operator size average $\Sigma_0(t)$. Since it is generically expected that operators grow linearly in time with the butterfly velocity $v_B$~\cite{Patel_butterfly_2017,Nahum_Operator_2018,Keyserlingk_Operator_2018}, the operator size average in closed Floquet systems is given by
\begin{equation}
\label{eq:ballistic}
  \Sigma_0(t)=c_0+2(t-1)v_B,
\end{equation}
where $c_0$ is a constant. For concreteness, we consider Floquet operators that are invariant under conjugation by arbitrary on-site unitaries (as for the DRPM below), in which case $c_0=1$.

Now, when dissipation with strength $p$ is turned on, we argue that the qualitative behavior of the operator size average changes depending on whether the effective dissipation strength $pt$ is weak ($pt\ll 1$) or strong ($pt\gg 1$). In the first case, since dissipation is sufficiently weak, we expect that the contributions from operator growth and bulk dissipation are effectively decoupled. As a result, we assert that the variation of the operator size average $\Delta\Sigma(t)=\Sigma(t+1)-\Sigma(t)$ satisfies
\begin{equation}\label{eq:op_size_weak_dissip}
    \Delta\Sigma(t)\simeq2v_B-c_{\mathrm{w}}p\Sigma_0(t),
\end{equation}
where $c_{\mathrm{w}}$ is a constant that depends on model parameters such as the interaction strength and the on-site Hilbert space dimension $q$. Note that we expect that the operator size average does not grow when there is no interaction, i.e., $\Delta\Sigma(t)=0$. 
Although Eq.~\eqref{eq:op_size_weak_dissip} is trivially solved, its most important aspect is that it indicates that dissipation does not influence the operator size average up to $t\lesssim p^{-1}$, which becomes a large timescale when dissipation is weak.

On the other hand, when dissipation is strong ($pt\gg 1$), the growth of the operator size average is dominated by bulk dissipation. In this case, the size does not grow, and even if it grows by one site, it is immediately dissipated. The growth of the size is thus controlled essentially by the single-site physics. To see how the variation of the operator size $\Delta\Sigma(t)$ depends on $p$ in this case, we note that, expanding $\mathcal{O}(t)=\sum_\alpha C_\alpha(t)P_\alpha$ where $C_\alpha(t)=\langle\mathcal{O}P_\alpha(t)\rangle$, a simple calculation (valid for large $pt$) gives
\begin{equation}
\Delta\Sigma(t)\simeq\frac{\sum_{\substack{\alpha:s_\alpha=1\\\beta:s_\beta=2}}|C_\alpha(t+1)|^2|C_\beta(t)|^2}{\sum_{\substack{\alpha:s_\alpha=1\\\beta:s_\beta=1}}|C_\alpha(t+1)|^2|C_\beta(t)|^2},
\end{equation}
where $s_\alpha=\langle P_\alpha\mathcal{S}\{P_\alpha\}\rangle$ is the size of the operator $P_\alpha$. Since it is generically expected that $|C_\alpha(t)|^2$, the probability of $P_\alpha$ evolving into $\mathcal{O}$ at time $t$, behaves as $|C_\alpha(t)|^2\sim \left.|C_\alpha(t)|^2\right|_{p=0}(1-p)^{2s_\alpha t}$, we conclude that
\begin{equation}\label{eq:op_size_strong_dissip}
    \Delta\Sigma(t)\simeq c_{\mathrm{s}}(1-p)^{2t},
\end{equation}
where $c_{\mathrm{s}}$ is another constant that in principle depends on model parameters. One important consequence of Eq.~\eqref{eq:op_size_strong_dissip} is that the operator size average saturates to a constant value
\begin{equation}
\lim_{t\to\infty}\Sigma(t)=\Sigma(t_0)+\frac{c_{\mathrm{s}}(1-p)^{2t_0}}{1-(1-p)^{-2}},
\end{equation}
where $t_0$ is the earliest time after which Eq.~\eqref{eq:op_size_strong_dissip} holds.
For the intermediate regime $pt\sim 1$, we expect that the competition between unitary operator growth and dissipation amounts to a nontrivial evolution equation, which generically depends on the details of the model and hence is not universal.

\subsection{Dynamics of Loschmidt echoes from operator growth}

Since for unitary dynamics the Loschmidt echo $\langle \mathcal{O}(t)\mathcal{O}(t)\rangle$  does not decay in time, for dissipative dynamics its decay is solely induced by dissipation. As in the operator size average, we expect that the behavior of the Loschmidt echo also changes qualitatively depending on whether $pt\ll 1$ or $pt\gg 1$.

We start by noting that the norm admits the following decomposition:
\begin{equation}\label{eq:decomp}
    \langle\mathcal{O}(t)\mathcal{O}(t)\rangle=\sum_{P_\alpha\in\mathcal{P}}\langle \mathcal{O}(t)P_\alpha\rangle^2,
\end{equation}
where we used the fact that $D^{-1}\sum_{P_\alpha\in\mathcal{P}}P_\alpha\otimes P_\alpha$ is the swap operator acting on the doubled Hilbert space (equivalently, by inserting a resolution of the identity in the doubled space). We can thus interpret the operator norm as the sum of the probabilities of the initial local operator $\mathcal{O}(0)$ evolving into the operator basis $P_\alpha$ at time $t$, which in turn implies that the operator norm is nothing but the probability of the operator $\mathcal{O}$ surviving up to time $t$ under dissipation.

When $pt\ll 1$, we postulate that dissipation affects this probability only via the operator size average $\Sigma(t)$, which we know remains unchanged in time in this case, i.e., $\Sigma(t)\simeq \Sigma_0(t)$. Since every nonidentity operator is dissipated with the survival probability $(1-p)^{2}$ at each time step, the survival probability for an operator string of size $\Sigma_0(t)$ after one time step is given by $(1-p)^{2\Sigma_0(t)}$. Therefore, the net survival probability of $\mathcal{O}(t)$ at time $t$ can be obtained by multiplying all the survival probabilities at every time step, which results in
\begin{equation}
\label{eq:t<<tp}
\begin{split}
    \langle\mathcal{O}(t)\mathcal{O}(t)\rangle
    &\simeq (1-p)^{2\sum_{t'=1}^{t-1}\Sigma_0(t')}
    \\
    &= (1-p)^{2(t-1)(1+v_B(t-2))}.
\end{split}
\end{equation}

Once $pt\gg 1$, the operator size eventually saturates. The decay of the Loschmidt echo is governed by the single-site physics because the longer the operator string $P_\alpha$ is, the less likely it is that the operator $\mathcal{O}$ evolves into it at time $t$. This means that the leading contribution to the decomposition Eq.~\eqref{eq:decomp} comes from the probability of $\mathcal{O}$ evolving in a way that conserves its support, i.e., that $P_\alpha$ is supported on the same site as $\mathcal{O}$. We expect that such a correlator  $\langle \mathcal{O}(t)P_\alpha\rangle^2$ decays as $((1-p)e^{-\lambda})^{2t}$ up to a numerical prefactor, where $2\lambda$ is the decay rate of $\langle \mathcal{O}(t)P_\alpha\rangle^2$ when dissipation is absent (i.e., for $p=0$). We thus conclude that the late time asymptotics of the norm is given by
\begin{equation}
\label{eq:t>>tp}
     \langle\mathcal{O}(t)\mathcal{O}(t)\rangle\sim\sum_{\mathrm{supp}P_\alpha=\mathrm{supp}\mathcal{O}}\langle \mathcal{O}(t)P_\alpha\rangle^2\sim ((1-p)e^{-\lambda})^{2t}.
\end{equation}

In summary, the late-time decay of the operator norm is distinguished by the time scale $t_p=1/p$ as
\begin{equation}\label{eq:norm_asymp}
    \langle\mathcal{O}(t)\mathcal{O}(t)\rangle\sim\begin{cases}
       (1-p)^{2(t-1)(1+v_B(t-2))}, & t\ll t_p\\
      (1-p)^{2t}e^{-2\lambda t}, & t\gg t_p
    \end{cases}.
\end{equation}
Note that during the first stage of the decay, there is generically a quadratic contribution in the exponential unless the intersite coupling is small. This behavior gradually changes into the exponential decay as $p$ becomes greater than $t^{-1}$. While such a change has been predicted to occur~\cite{jalabert2001,jacquod2001,goussev2012arXiv}, the critical dissipation rate across which the behavior alters has always been assumed to be independent of time. Here, we showed that this is generically not the case and that it depends on time nontrivially.

\section{Proof for an exactly solvable model}

\subsection{Loschmidt echo in the DRPM}

We now provide an analytical proof of Eq.~\eqref{eq:norm_asymp} for the DRPM~\cite{yoshimura2024,yoshimura2025}, which is a prototypical exactly-solvable Floquet many-body circuit subjected to bulk dissipation. 
Let us first recall the definition of the unitary random phase model (RPM)~\cite{Chan_Spectral_2018}. The Floquet operator of the RPM is $W=W_2W_1$, where $W_1=U_1\otimes\dots\otimes U_L$ consists of $q\times q$ Haar unitaries $U_x$, whereas $W_2$ couples neighboring sites and acts diagonally on the computational basis $|a_1,a_2,\dots,a_L\rangle\in(\mathbb{C}^q)^L=\mathcal{H}$, where $a_x=0,\dots,q-1$, with phase $\exp\left(\ii\sum_{x=1}^L\varphi_{a_x,a_{x+1}}\right)$. Each $\varphi_{a_x,a_{x+1}}$ is independently Gaussian distributed with mean zero and variance $\varepsilon>0$.
In the DRPM, each Floquet step is accompanied by a quantum channel that introduces dissipation. In this paper, we assume that the quantum channel acts on each site independently, $\Phi=\bigotimes_x\Phi_x$. To make contact with the previous sections, we focus again on the depolarizing channel, $\Phi_x(\varrho_x)=(1-p)\varrho_x+p I_q/q$, although the results hold for arbitrary on-site quantum channels~\cite{yoshimura2024,yoshimura2025}.
The DRPM becomes exactly solvable in the large-$q$ limit after averaging over the Haar-random unitary gates, which we denote by an overline, $\overline{(\cdots)}$. 
To compute the operator norm $\overline{\langle\mathcal{O}(t)\mathcal{O}(t)\rangle}$, we first briefly recall the diagrammatic technique, which was originally developed for evaluating the OTOC in the large-$q$ RPM in Ref.~\cite{yoshimura2023operator}. 

It was observed in Ref.~\cite{yoshimura2023operator} that, upon Haar-averaging and the large-$q$ limit, the size of the transfer matrix for objects that contain two copies of the Haar unitary $U_i$ and its conjugate $U_i^*$---e.g., the OTOC and the norm---reduces from $q^{4t}$ to $2t+1$. Such ($2t+1$)-dimensional transfer matrix $S$ acts on the ($2t+1$)-dimensional vector space spanned by the leading $s=0,\dots,2t$ pairings of the indices of the Haar-random unitaries,
\begin{equation}
    S=\begin{pmatrix}
        S_1 & S_2 \\
        S_2^\mathrm{T} & S_3
    \end{pmatrix},
\end{equation}
where $S_1$, $S_2$, and $S_3$ are $(t+1)\times(t+1)$, $(t+1)\times t$, and $t\times t$ matrices, respectively. Their matrix elements are given by $[S_1]_{ab}=\delta_{ab}+\rho^{|a-b|-1}(1-\delta_{ab})$, $[S_2]_{ab}=\delta_{ab}+\rho^{|a-b|-1}(1-\delta_{ab}+\Theta(b-a)(\rho-1))$, and $[S_3]_{ab}=\delta_{ab}+\rho^{|a-b|}(1-\delta_{ab})$, where indices $a,b$ run from $0$ to $t$, $\Theta(a)$ is the step function with $\Theta(0)=0$, and $\rho=e^{-2\varepsilon}$. The presence of dissipation manifests itself in a diagonal on-site matrix $E$ of size $2t+1$, namely, $E_\mathrm{d}=\mathrm{diag}(\omega^{t-1},\omega^{t-2},\dots,1,1,-\omega^{t-1},\dots,-1)$, 
where $\omega=(1-p)^2$.

We recently showed in Ref.~\cite{yoshimura2025} that, in terms of these matrices, the operator norm $\overline{\langle\mathcal{O}(t)\mathcal{O}(t)\rangle}$ is succinctly expressed as $\overline{\langle\mathcal{O}(t)\mathcal{O}(t)\rangle}=[\hat{S}^{L-1}_\mathrm{d}]_{00}$ where $\hat{S}_\mathrm{d}=E_\mathrm{d}S$ when the system is subject to periodic boundary conditions. 
In Ref.~\cite{yoshimura2025}, in order to study anomalous relaxation~\cite{sa2022PRR,garcia2023PRD2,scheurer2023ARXIV,mori2024,yoshimura2024} and the related RP resonances~\cite{prosen2002,prosen2004,prosen2007,mori2024,yoshimura2024,znidaric2024,znidaric2024b,jacoby2024,zhang2024,yoshimura2025,znidaric2025,duh2025}, we focused on the case where we take the large-$t$ limit before the thermodynamic limit; here, we consider the opposite order. 

Importantly, the transfer matrix $\hat{S}_\mathrm{d}$ becomes independent of $L$ for $L>2(t-1)$~\cite{yoshimura2023operator,yoshimura2025}. The operator norm in the thermodynamic limit, $\mathcal{N}(t):=\lim_{L\to\infty} \overline{\langle\mathcal{O}(t)\mathcal{O}(t)\rangle}$, is thus given by~\cite{yoshimura2025}
\begin{equation}
\label{eq:qdeformation_OO}
   \mathcal{N}(t)
    =(\omega\rho^2)^{(t-1)}\left[\left(1-\rho^{-1};\omega\right)_{t-1}\right]^2,
\end{equation}
where $(a;q)_n=\prod_{k=0}^{n-1}(1-aq^k)$ is the $q$-Pochhammer symbol. Note that the result holds for any boundary conditions as the thermodynamic limit is taken first now. From Eq.~(\ref{eq:qdeformation_OO}), we see that the late-time behavior of the operator norm changes qualitatively over the timescale $t_p=1/p$. First, when $t\ll t_p$ we can approximate $\log(1-(1-\rho^{-1})\omega^k)\simeq -\log\rho-2p(1-\rho)k$, which gives
\begin{align}
    &\log(1-\rho^{-1};\omega)_{t-1}\n
    &\simeq -(t-1)\log\rho-2p(1-\rho)\sum_{k=0}^{t-2}k\n
    &=-(t-1)\log\rho+\log(1-p)v_B(t-1)(t-2),
\end{align}
where we defined the butterfly velocity $v_B=1-\rho$. In this regime the operator norm thus behaves as $\overline{\langle\mathcal{O}(t)\mathcal{O}(t)\rangle}\simeq (1-p)^{2(t-1)(1+v_B(t-2))}$, in agreement with Eq.~\eqref{eq:t<<tp}. On the other hand, when $t\gg t_p$, the $q$-Pochhammer symbol converges to a constant value $\left(1-\rho^{-1};\omega\right)_{\infty}$ and therefore the decay rate is given simply by the prefactor $(\omega \rho^2)^{(t-1)}$, where we used that $t\gg t_p$, in agreement with Eq.~\eqref{eq:t>>tp} if we identify $\lambda=2\varep$.

\subsection{Operator spreading and OTOCs in the DRPM}

Before proceeding to characterizing operator spreading in the DRPM, let us first compute the average operator size in the unitary RPM. In the RPM, the norm $\langle \mathcal{O}(t)\mathcal{O}(t)\rangle$ is conserved, $\langle \mathcal{O}(t)\mathcal{O}(t)\rangle=1$, and the average operator size, which we denote by $\overline{\Sigma_0(t)}$, can be expressed as $\overline{\Sigma_0(t)} =\sum_x\overline{\mathcal{C}(x,t)}$. Using the exact OTOC at large $q$ for infinite volume $L\to\infty$ \cite{yoshimura2023operator},
\begin{equation}\label{eq:otoc_nospin}
    \lim_{L\to\infty}\overline{\mathcal{C}(x,t)}=
        \sum_{i=0}^{t-|x|-1}
        \begin{pmatrix}
        t-1\\
        i
        \end{pmatrix}
        \rho^i(1-\rho)^{t-i-1},
\end{equation}
we readily obtain
\begin{equation}
\label{eq:ballistic_2}
    \overline{\Sigma_0(t)}=1+2(t-1)v_B.
\end{equation}

Now, for dissipative random circuits, taking a disorder average of Eq.~\eqref{eq:OTOC_size} with respect to random unitaries poses an immediate problem, since not only the numerator but also the denominator contains those unitaries. However, it turns out that for the DRPM at large $q$, Haar averaging can be performed independently (see Appendix~\ref{app:averages} for the derivation). Furthermore, the Haar-averaged OTOC does not depend on the choice of $P_{\alpha^x}$ in the DRPM, similarly to the RPM~\cite{yoshimura2023operator}. The average operator size in the $q\to\infty$ limit is therefore given by
\begin{equation}\label{eq:operator_size2}
    \overline{\Sigma(t)}
    =L-\frac{\sum_x\overline{\langle \mathcal{O}(t)P_{\alpha^x}\mathcal{O}(t)P_{\alpha^x}\rangle}}{\overline{\langle \mathcal{O}(t)\mathcal{O}(t)\rangle}}.
\end{equation}

As for the operator norm, the average operator size $\overline{\Sigma(t)}$ can be evaluated diagramatically: for periodic boundary conditions, its four-point contribution reads
\begin{equation}
\label{eq:OTOC_OPOP}
    \overline{\langle \mathcal{O}(t)P_{\alpha^x}\mathcal{O}(t)P_{\alpha^x}\rangle}=[\hat{S}^{L-x}_\mathrm{d}]_{t0}[\hat{S}^{x}_\mathrm{d}]_{0t},
\end{equation}
which allows us to write the average operator size as
\begin{equation}\label{eq:drpm_opsize1}
    \overline{\Sigma(t)}=L-\frac{1}{\langle\mathcal{O}(t)\mathcal{O}(t)\rangle}\sum_{x=1}^L[\hat{S}^{L-x}_\mathrm{d}]_{t0}[\hat{S}^{x}_\mathrm{d}]_{0t}.
\end{equation}
When $L>2(t-1)$ the operator size becomes independent of $L$ as the time-evolved operator stays within the light cone, and as a result Eq.~\eqref{eq:drpm_opsize1} reads
\begin{equation}\label{eq:drpm_opsize2}
   \overline{\Sigma(t)}=2t-1-\frac{2}{\mathcal{M}(t)}\sum_{x=1}^{t-1}[\hat{S}^x_\mathrm{d}]_{t0},
\end{equation}
where $\mathcal{M}(t)=[\hat{S}^{L>t-1}_\mathrm{d}]_{t0}=\rho^{t-1}(1-\rho^{-1},\omega)_{t-1}$. It turns out that the sum can be carried out explicitly, yielding the following closed expression of the operator size average:
\begin{equation}
    \overline{\Sigma(t)}=2t-1-2\sum_{n=0}^{t-2} \frac{1}{1-(1-\rho^{-1})\omega^n}.
\end{equation}
As expected, this result reduces to Eq.~(\ref{eq:ballistic}) exactly in the dissipationless limit $p=0$.

To see how the operator size average grows, it is again convenient to study its variation
\begin{equation}
    \overline{\Delta\Sigma(t)}=\frac{2(1-\rho^{-1})\omega^{t-1}}{1-(1-\rho^{-1})\omega^{t-1}},
\end{equation}
which behaves qualitatively differently depending on the value of $pt$:
\begin{equation}
    \overline{\Delta\Sigma(t)}\simeq\begin{cases}
        2v_B-2\rho p\overline{\Sigma_0(t)}, & pt\ll1\\
       2(\rho^{-1}-1)(1-p)^{2t}, & pt\gg1
    \end{cases}.
\end{equation}
This shows that the heuristic variation equations Eqs.~\eqref{eq:op_size_weak_dissip} and \eqref{eq:op_size_strong_dissip} we introduced earlier are indeed satisfied in the DRPM with $c_\mathrm{w}=2\rho$ and $c_\mathrm{s}=2(\rho^{-1}-1)$.

\section{Conclusions}

We provided a general picture of how operator growth is affected by dissipation in open locally-interacting quantum many-body systems, and verified it by analytically working out the dissipative random phase model (DRPM), a minimal open Floquet circuit. We found that the decay of the Loschmidt echo changes qualitatively as the effective perturbation strength $pt$, which is time dependent, is varied. This means that the behavior of the Loschmidt echo can also change in time, which, to our knowledge, has been overlooked in the literature.
We also demonstrated that the operator size average saturates to a finite, nonzero value at late times regardless of the dissipation strength in the DRPM, which we expect also happens in generic locally-interacting open Floquet systems, contrary to the findings of Ref.~\cite{schuster2023PRL}. We conjecture that the saturation was not seen there because of the limits to the timescale they had access to numerically.

Having established the general behavior of operator growth in locally-interacting systems, it is natural to wonder what would happen to systems that are non-locally interacting, e.g., all-to-all quantum circuits. While a heuristic picture for these systems was also proposed in Ref.~\cite{schuster2023PRL}, it would be highly desirable to corroborate it by means of exact results, extending the formalism employed in this paper.

\begin{acknowledgments}
We thank Curt von Keyserlingk for useful discussions.
L.S.\ was supported by a Research Fellowship from the Royal Commission for the Exhibition of 1851.
\end{acknowledgments}

\renewcommand{\appendixname}{APPENDIX}
\appendix

\section{\uppercase{Equivalence of annealed and quenched averages in the large}-$q$ DRPM}
\label{app:averages}

Here, we show that Haar averaging of the denominator and the numerator of the operator size average can be taken independently in the large-$q$ DRPM. Namely, we demonstrate
\begin{equation}\label{eq:ave_indep}
    \overline{ \frac{\mathcal{C}(x,t)}{\langle \mathcal{O}(t)\mathcal{O}(t)\rangle}}=\frac{\overline{\mathcal{C}(x,t)}}{\overline{\langle \mathcal{O}(t)\mathcal{O}(t)\rangle}}
\end{equation}
in the large-$q$ limit. To see this, we first note
\begin{equation}\label{eq:summand}
    \frac{\mathcal{C}(x,t)}{\langle \mathcal{O}(t)\mathcal{O}(t)\rangle}=\mathcal{C}(x,t)\sum_{k=0}^\infty(1-\langle \mathcal{O}(t)\mathcal{O}(t)\rangle)^k,
\end{equation}
where we used $0\leq\langle \mathcal{O}(t)\mathcal{O}(t)\rangle\leq \langle \mathcal{O}(0)\mathcal{O}(0)\rangle=1$. Since the OTOC is given by Eq.~\eqref{eq:def:OTO2}, it therefore suffices to show that $\overline{\langle \mathcal{O}(t)\mathcal{O}(t)\rangle^m}=(\overline{\langle \mathcal{O}(t)\mathcal{O}(t)\rangle})^m$ holds at large $q$. This follows easily: to illustrate the idea, let us consider the case $t=1$ and $m=2$ where we have two replicas of a diagram that contains two pairs of a unitary $U_i$ and its conjugate $U_i^*$. Haar averaging then contracts every unitary with one of the conjugates, and it can be readily observed that when a unitary in one diagram is paired with a conjugate in another diagram, this induces an extra factor of $q^{-2}$ as more legs of the diagrams are now labeled by the same index. This observation persists also for generic $t$ and $m$, indicating that we generate the minimum number of $q^{-1}$ by contracting unitaries within the diagram they constitute. We thus conclude that $\overline{\langle \mathcal{O}(t)\mathcal{O}(t)\rangle^m}=(\overline{\langle \mathcal{O}(t)\mathcal{O}(t)\rangle})^m$ holds, and as a result, Eq.~\eqref{eq:ave_indep} is true at large $q$.
Note that the mechanism here is essentially similar to how the minimal cut configuration dominates in the ensemble-averaged second R\'enyi entropy $\overline{S_2(t)}$ in the random unitary circuit at large $q$.

\bibliography{bib.bib}

\end{document}